\documentclass[conference]{IEEEtran}
\IEEEoverridecommandlockouts
\usepackage{cite}
\usepackage{amsmath,amssymb,amsfonts}
\usepackage{algorithmic}
\usepackage{graphicx}
\usepackage{textcomp}
\usepackage{xcolor}
\usepackage{multirow}
\usepackage{tabularx}
\usepackage{booktabs}
\usepackage[utf8]{inputenc} 
\usepackage[T1]{fontenc}    
\usepackage{hyperref}       
\usepackage{url}            
\usepackage{nicefrac}       
\usepackage{microtype}      
\usepackage[mathscr]{eucal}
\usepackage{algorithm}
\usepackage{algorithmic}
\usepackage{array,makecell}
\usepackage{wrapfig}
\usepackage{float}
\usepackage{overpic}
\usepackage{caption}
\usepackage{subfigure}
\usepackage{bbding}
\usepackage{enumitem}

\usepackage[misc]{ifsym}

\def\BibTeX{{\rm B\kern-.05em{\sc i\kern-.025em b}\kern-.08em
    T\kern-.1667em\lower.7ex\hbox{E}\kern-.125emX}}
\begin{document}

\title{EAD-VC: Enhancing Speech Auto-Disentanglement for Voice Conversion with IFUB Estimator and Joint Text-Guided Consistent Learning}

\author{\IEEEauthorblockN{Ziqi Liang$^{1,2\ddagger}$, Jianzong Wang$^{1\ddagger}$\thanks{$\ddagger$ Equal Contributions}, Xulong Zhang$^{1\textrm{\Letter}}$\thanks{$^\textrm{\Letter}$Corresponding author: Xulong Zhang (zhangxulong@ieee.org).}, Yong Zhang$^{1}$, Ning Cheng$^{1}$, Jing Xiao$^{1}$}
\IEEEauthorblockA{\textit{$^{1}$Ping An Technology (Shenzhen) Co., Ltd.}\\\textit{$^{2}$University of Science and Technology of China}}
}


\maketitle

\begin{abstract}
Using unsupervised learning to disentangle speech into content, rhythm, pitch, and timbre for voice conversion has become a hot research topic. 
Existing works generally take into account disentangling speech components through human-crafted bottleneck features which can not achieve sufficient information disentangling, while pitch and rhythm may still be mixed together. There is a risk of information overlap in the disentangling process which results in less speech naturalness. 
To overcome such limits, we propose a two-stage model to disentangle speech representations in a self-supervised manner without a human-crafted bottleneck design, which uses the Mutual Information (MI) with the designed upper bound estimator (IFUB) to separate overlapping information between speech components.
Moreover, we design a Joint Text-Guided Consistent (TGC) module to guide the extraction of speech content and eliminate timbre leakage issues. 
Experiments show that our model can achieve a better performance than the baseline, regarding disentanglement effectiveness, speech naturalness, and similarity. 
Audio samples can be found at \href{https://largeaudiomodel.com/eadvc}{https://largeaudiomodel.com/eadvc}.
\end{abstract}


\section{Introduction}
Voice Conversion (VC) involves transforming the vocal characteristics of an source speaker into those of a target speaker. This is achieved by altering the speech para-linguistic aspects (e.g. speaker identity, prosody), without compromising the original linguistic information. The maturity of VC has brought benefits to various industries\cite{asr_voice_attack, asr_box_attack, assisent}.

With the advancement of deep learning, various voice conversion solutions have been proposed. Some approaches utilize auxiliary models such as ASR or TTS models to achieve VC \cite{mixvc, Expressivevc, CE-CTC-BN}. In \cite{StarGAN-VC,cycleGAN-VC,VAEVC1,VAEVC2,tang2021TGAVC,zhang2021CycleGEAN,tang2022AVQVC}, researchers use GAN and VAE to produce speech resembling the target speakers, but training GAN-based models is typically challenging.
Recently, a lot of VC systems have enabled speech representation of speaker-dependent and independent information \cite{autovc, Multi-target-VC, oneshotvc, AVQVC}. 
These works decompose speech into speaker and content representations, ensuring not only distribution matching like GANs but also easy training as easily as VAEs.
\cite{AVQVC} proposed a method for disentangling content and speaker information from speech. They used a vector quantization-based method to eliminate speaker information in the content information, and then add the unseen speaker information in the decoding stage, which greatly improves the model generalization for unknown speakers.
\cite{Qian2020F0ConsistentMN} proposed to unify other components besides timbre and content into prosodic features, and proposed a VC model that uses F0 as a condition. 
But these methods can only decompose speech into speaker, content and prosody that achieve coarse-grained disentanglement, while the information of pitch and rhythm are still mixed together. SpeechFlow \cite{speechsplit1} uses multiple autoencoders to disentangle speech into four components of pitch, rhythm, timbre, and content by introducing three well-designed information Bottlenecks. SpeechSplit2.0 \cite{speechsplit2} relies on the architecture provided by SpeechFlow. By employing additional signal processing techniques, the speech can be disentangled without the need for laborious bottleneck tuning. 

However, the model should be capable of independently distinguishing speech, thereby eliminating the need for manual extraction of bottleneck features. It not only conserves time but also minimizes the potential for bias and subjectivity that can stem from the manual selection of features.
Liu et al.\cite{auto-disentangle} realize that disentangling speech representation to the content, pitch, and rhythm by comparing the speech and its augmented version through ranking, which does not require bottleneck fine-tuning. However, the decoupling method based on data augmentation also has the same problem of insufficient decoupling, causing content, pitch, and rhythm to still be entangled. 
Moreover, without the guidance of linguistic information, the embedding extracted by the content encoder may be mixed with speaker information, causing timbre leakage and content inconsistencies.

\begin{figure*}[htp]
    \centering
    \includegraphics[width=16.5cm]{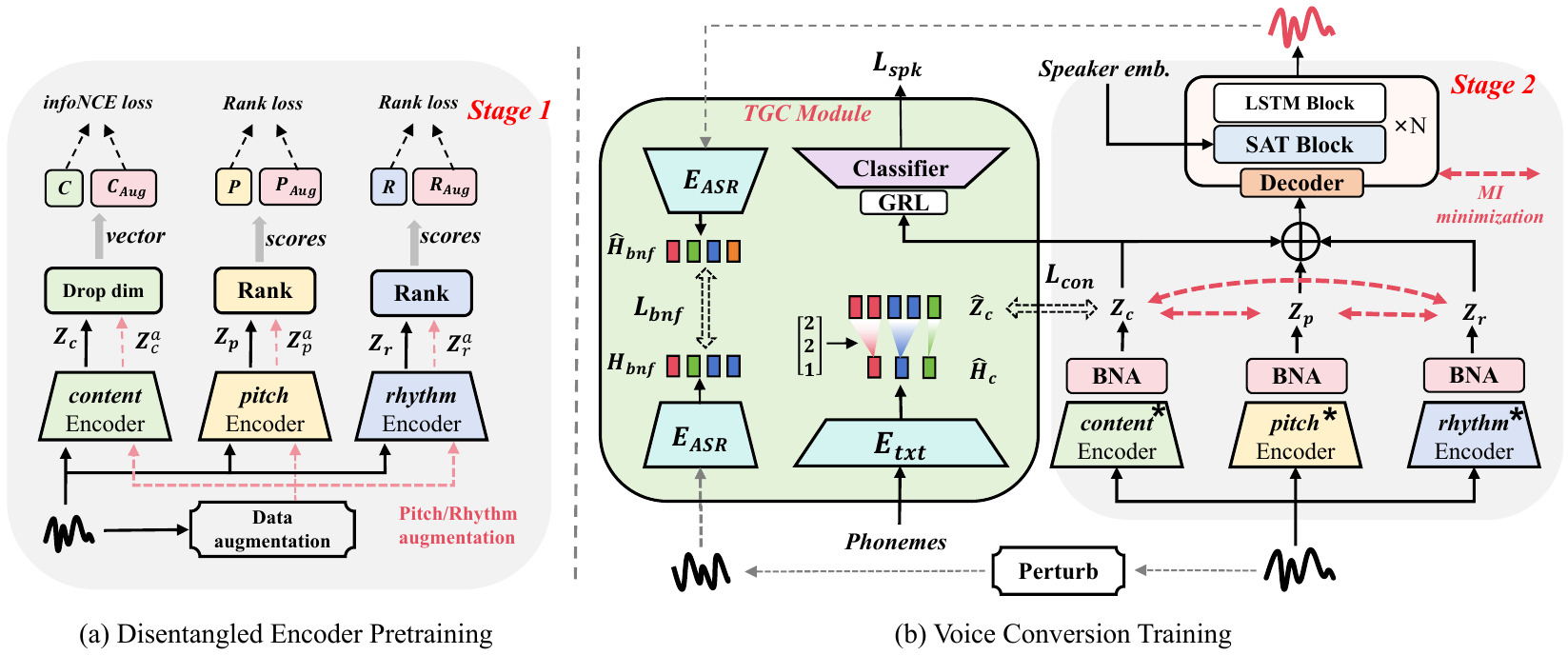}
    \caption{Framework of EAD-VC, which shows the two stages of our method: ({\uppercase\expandafter{\romannumeral1}}) Train the encoder based on the data and its augmented versions to disentangle speech as (a).
    ({\uppercase\expandafter{\romannumeral2}}) Freeze encoders to extract $Z_{c}$, $Z_{p}$, and $Z_{r}$ in (b); Desired content embedding $\hat{Z_{c}}$ from phonemes, which is used to guide the content encoder training. $E_{ASR}$ is used to keep the content consistent after VC.} 
    \vspace{-0.5cm}
    \label{stage2}
\end{figure*}


To obtain the converted speech without human-crafted bottleneck tuning, we turn to the decoupling method based on self-supervised learning and focus on eliminating information overlap between different speech components. Thus, we propose the two-stage VC model for \textbf{E}nhanced speech \textbf{A}uto-\textbf{D}isentanglement (EAD-VC). 
The primary contributions of our work are as follows:
\begin{itemize}
\item Disentangling speech representations into content, pitch, rhythm, and timbre in a self-supervised manner without human-crafted bottleneck tuning.
\item We design a new upper bound estimator IFUB of MI to enhance the decoupling between content, pitch, rhythm, and timbre. The bottleneck adapter (BNA) trained with MI with IFUB is designed to separate overlapping information between different speech components.
\item We propose a joint text-guided consisent (TGC) module to solve timbre leakage in the content extraction and avoid content inconsistencies after conversion under the guidance of text transcriptions and ASR-Bottleneck (ASR-BNFs).
\end{itemize}

\section{Related work}
With the advancement of deep learning, many voice conversion has explored many techniques employing VAEs or GANs to facilitate the transfer of speaker information.
VAE-VC \cite{vaevc} accomplishes voice conversion by generating speaker-independent content embeddings through its encoder. CDVAE-VC \cite{cdvaevc} uses two VAEs to reconstruct two different speech features of straight spectra and mel-cepstral coefficients (MCCs) respectively. ACE-VC \cite{ACVAEVC} incorporates an auxiliary speaker classifier after the decoder and prevents the classifier from correctly classifying speaker information. Influenced by style transfer techniques in computer vision, \cite{cycleGAN-VC} and \cite{StarGAN-VC} introduced CycleGAN and StarGAN respectively to implement voice conversion. Later, some methods utilize TTS or ASR models as auxiliary models to implement voice conversion \cite{Mix-Guided-VC,Expressive-VC,Zhao2022DisentanglingCA}. 
Some researchers have harnessed the power of GANs and VAEs to produce voice that closely resemble the voice of a target speaker.~\cite{Kameoka2018StarGANVCNM, Fang2018HighQualityNV, zhang2021CycleGEAN,zhang2023Voice}, but GAN-based voice conversion models are usually difficult to train. 
Recently, many VC systems have achieved the decoupling of speaker representation and speech-independent representation (content representation) \cite{Adain-vc,autovc,deng2023PMVC,deng2024Learning,deng2024ctvc,PolyakACKLHMD21}. Compared with traditional voice conversion, it can achieve no need for the paired source speaker and target speaker data during training. 
High-quality source and target speech inputs significantly enhance the performance of the disentangled-based VC method, leading to substantial advancements in both the fidelity and the resemblance.
Current mainstream VC models usually focus on speech representation disentanglement, aiming to disentangle speaker information and content information as much as possible \cite{autovc,Adain-vc,Lu2019OneShotVC}. AutoVC \cite{autovc} combines the ideas of GAN \cite{Fang2018HighQualityNV} and CVAE \cite{ACVAEVC} to decouple content information and speaker information, and achieves One-Shot voice conversion. At the same time, there are also some works that pay more attention to the prosody of speech. \cite{Qian2020F0ConsistentMN} unifies components other than timbre and content into prosody features, and proposes a VC model using F0 as a condition, which adjusts the Auto-Encoder through constraining bottleneck features to achieve prosody decoupling.

If the model simply focus on timbre and content but ignores other components, it may result in less natural and expressive generated speech. Therefore, \cite{IQDUBBING} additionally extracts the rhythm and pitch components from speech to achieve more fine-grained information decoupling. \cite{speechsplit1,speechsplit2,Yang2022SpeechRD} use multiple encoders and signal processing techniques to disentangle speech into pitch, rhythm, content, and timbre by introducing three well-designed information bottlenecks.
Based on SpeechFlow \cite{speechsplit1}, our method explores the use of self-supervised learning to decouple content, pitch, rhythm, and timbre information without manually extracting pitch and rhythm information as guidance. At the same time, we enhance the information constraints between decoupled components and alleviate the information leakage problem existing in the above work.

\section{Methodology}

\subsection{SSL-based speech disentanglement}
As in \cite{speechsplit1, speechsplit2}, we employ three encoders in our model. Rank loss and contrastive learning are applied to extract pitch, rhythm, and content representations of speech in a self-supervised manner.
Pitch shift and time stretch are applied to modify the speech pitch and rhythm. As shown in Fig.\ref{stage2}(a), speech data $y$ and augmented data $y_{Aug}$ were send to the disentangled encoders. 
We get $Z_{c}$, $Z_{p}$ and $Z_{r}$ from $y$ with the three disentangled encoders, and $Z_{c}^{a}$, $Z_{p}^{a}$ and $Z_{r}^{a}$ from $y_{Aug}$. Afterward, we apply the Rank layer to map $Z_{r}$ and $Z_{p}$ into two individual scores $R$ and $P$, which is inspired by \cite{auto-disentangle, zeroshotvc, rank_layer}. 
Hyperparameter $\gamma\in(0,1)$ is used to indicate the data augmentation intensity. $\gamma<$ 0.5 means negative augmentations such as decreasing pitch or rhythm, while $\gamma>$ 0.5 means positive augmentations such as increasing pitch or rhythm. $\gamma$ = 0.5 indicates no augmentation is applied. To ensure that the encoders produce disentangled representations by recognizing this augmentation intensity, we first apply a sigmoid function on scores of pitch and rhythm:
\begin{equation}
        \small
        \mathcal{S}_{r} = \frac{1}{1+e^{-(R-R_{Aug})}} \quad
        \mathcal{S}_{p} = \frac{1}{1+e^{-(P-P_{Aug})}}
\end{equation}
then we get the rank loss based on rhythm scores $S_{r}$ and pitch scores $S_{p}$:
\begin{equation}
        \small
        \mathcal{L}_{r} = -\gamma^{r}log(S_{r})-(1-\gamma^{r})log(1-S_{r}) 
\end{equation}
\begin{equation}
        \small
        \mathcal{L}_{p} = -\gamma^{p}log(S_{p})-(1-\gamma^{p})log(1-S_{p})
\end{equation}

Concurrently, the content is initially condensed into a vector $C$ for subsequent processing. Furthermore, to ensure that the content encoder solely produces representations pertinent to the content, we implement a contrastive learning loss on both the original content $C$ and its augmented counterpart $C_{Aug}$:

\begin{footnotesize}
\begin{equation}
        \mathcal{L}_{infoNCE} = -\beta \cdot log\frac{sim(C,C_{Aug})}{sim(C,C_{Aug})+\sum_{x_{Neg}}{sim(C,C_{Neg})} }
        \label{nce_tem}
\end{equation}
\end{footnotesize}

\normalsize
\noindent where $sim(\cdot,\cdot)$ is the exponential dot product, with a temperature $t$. $\beta$ is a decay coefficient, the initial value is 1.0. Ultimately, we initiate the pre-training process for the trio of disentangled encoders utilizing the loss:
\begin{equation}
        \mathop{\min}_{E_{c}(\cdot),E_{p}(\cdot),E_{r}(\cdot)}\mathcal{L}_{enc} = L_{r}+L_{p}+ L_{infoNCE}
\end{equation}

\subsection{Mutual information with IFUB estimator}
Since full fine-tuning could distort pre-trained features and lead to worse performance in the presence of large distribution shifts, we freeze the pre-trained encoder in Fig.\ref{stage2} (a) to preserve model decoupling capabilities. We design a trainable bottleneck adaptor (BNA), and add it to each frozen encoder, which was trained to achieve further disentangling. 
Moreover, we designed a new mutual information upper bound estimator IFUB combined with InfoNCE, which inspired by vCLUB \cite{Cheng2020CLUBAC}. We utilize BNA trained with IFUB of MI to eliminate information overlap between pitch, rhythm, and content and enhance the decoupling of speech components. During training, the gradient only updates BNA layers while retaining the pre-trained parameters from frozen encoders. We remove disentangled information overlap and enhance disentanglement by MI minimization. 



While $Q_{\theta}(Y|X)$ in vCLUB \cite{Cheng2020CLUBAC,vqmivc} can be represented by any neural network, it is common in practice to parameterize $Q_{\theta}(Y|X)$ using a Gaussian family. The potential reason for avoiding the use of MLP to parameterize $Q_{\theta}(Y|X)$ is that it is difficult to converge in CLUB, so it makes sense to design a function that can parameterize $Q_{\theta}(Y|X)$ with any neural network and converge easily. 
Therefore, we integrate vCLUB with InfoNCE to to design a new upper bound estimator IFUB. This integration involves initially using the trained $f(x_{i},y_{i})$ from InfoNCE to substitute $\log p(Y|X)$ in vCLUB for computing the value of $\mathcal{\hat{I}}(X,Y)$, and subsequently minimizing it to reduce MI.
We depict the process of minimizing Mutual Information using IFUB in Algorithm 1. 

\begin{algorithm}[h]
\caption{MI minimization with IFUB estimator}
\label{mi}
\textbf{Training}: 
\begin{algorithmic}
\FOR{each iteration} 
\STATE {Sample $\{ x_{i},y_{i} \}_{i=1}^{N}$ from $E_{c,p,r}(x,y)$}.
\STATE {Update Critic $f_{c,p,r}(\cdot)$ by maximizing $\mathcal{L}(x_{i},y_{i})$}:
\STATE {$\mathcal{L}(x_{i},y_{i}) = \frac{1}{N}\sum_{i=1}^{N} \log \frac{exp(f_{c,p,r}(x_{i},y_{i}))}{\frac{1}{N} \sum_{j=1}^{N} exp(f_{c,p,r}(x_{i},y_{j}))} $}.
\FOR{$i=0$; $i<N$; $i++$}
\STATE {$UB_{c,p,r}^{i} = f_{c,p,r}(x_{i},y_{i}) - \frac{1}{N}\sum_{j=1}^{N}f_{c,p,r}(x_{i},x_{j})$}.
\ENDFOR
\STATE {Update $\textbf{E}_{c}$ and $\textbf{E}_{p}$ by }:
\STATE {$\mathcal{\hat{I}}(Z_{c}, Z_{p}) = \frac{1}{N}\sum_{i=1}^{N}UB_{c,p}^{i}$}.
\STATE {Update $\textbf{E}_{p}$ and $\textbf{E}_{r}$ by }:
\STATE {$\mathcal{\hat{I}}(Z_{p}, Z_{r}) = \frac{1}{N}\sum_{i=1}^{N}UB_{p,r}^{i}$}.
\STATE {Update $\textbf{E}_{p}$ and $\textbf{E}_{r}$ by }:
\STATE {$\mathcal{\hat{I}}(Z_{c}, Z_{r}) = \frac{1}{N}\sum_{i=1}^{N}UB_{c,r}^{i}$}.
\ENDFOR
\end{algorithmic}
\end{algorithm}

During the training phase, the neural network $f_{c,p,r}(\cdot)$ and $E_{c,p,r}(\cdot)$ are trained alternately, and we parameterize $f_{c,p,r}(\cdot)$ using a fully connected layer. By minimizing $\mathcal{L}_{MI}$, we can reduce the correlation among various speaker-irrelevant speech representations (content, pitch, and rhythm) and realize enhanced speech auto-disentanglement. 
The estimated MI loss is as:
\begin{equation}
        \mathop{\min}_{Ada(\cdot)}\mathcal{L}_{MI} = \mathcal{\hat{I}}(Z_{c}, Z_{p}) + \mathcal{\hat{I}}(Z_{p}, Z_{r}) + \mathcal{\hat{I}}(Z_{c}, Z_{r})
\end{equation}


\subsection{Joint text-guided consistent learning}
We propose a TGC module to solve the problem of timbre leakage in the content encoder, which contains four submodules: ({\uppercase\expandafter{\romannumeral1}}) Text2Content module with length regulator. ({\uppercase\expandafter{\romannumeral2}}) Shared Speech2Content module trained with CTC loss. ({\uppercase\expandafter{\romannumeral3}}) Adversarial speaker classifier. ({\uppercase\expandafter{\romannumeral4}}) Timbre fusion.

\textbf{Text2Content module}:
To guide the content encoder in generating speaker-independent content embedding, we utilize a text encoder that produces text embedding $\hat{Z_{c}}$ from phonemes.
Then we calculate content consistent loss: 
\begin{equation}
        Z_{c} = E_{Ada}(E_{c}(x_{i}^{sp}))
\end{equation}
\begin{equation}
        \hat{Z}_{c} = F_{dur}(E_{txt}(x_{i}^{phn}))
\end{equation}
\begin{equation}
    \small
    \begin{split}
        \mathop{\min}_{E_{c}(\cdot)Ada(\cdot)}\mathcal{L}_{con} = \frac{1}{N}\sum_{i=1}^{N}(\parallel Z_{c}-\hat{Z}_{c}) \parallel_{1})  \\
    \end{split}
\end{equation}
\noindent where $x_{i}^{sp}$ represents mel-spectrogram and $x_{i}^{phn}$ represents phonemes. $x_{i}^{sp}$ passes through the content encoder $E_{c}$ and BNA layer $E_{Ada}$ in turn to get content embedding $Z_{c}$. $x_{i}^{phn}$ passes through the text encoder $E_{txt}$ and duration predictor $F_{dur}$ to get text2content embedding $\hat{Z}_{c}$, which is obtained by phoneme alignment of $\hat{H}_{c}$ according to the duration.   
The aim of $L_{con}$ is to encourage the content encoder to produce speaker-independent content embedding from source speech with the guidance of text embedding from phonemes.

\textbf{Speech2Content module}: 
We use the shared ASR encoder $E_{asr}$ \cite{asr_ctc} trained by CTC Loss to extract the ASR bottleneck features (ASR-BNFs) from converted audio and source audio with formant perturbing. ASR-BNFs is the bottleneck feature extracted from the penultimate layer of the ASR model trained based on CTC loss. It only contains linguistic information.
We first perform a formant perturb $DA_{per}$ on the $x_{i}^{sp}$ and eliminate the timbre information to obtain $\tilde{x}_{i}^{sp}$. 
Both the perturbed audio $\tilde{x}_{i}^{sp}$ and the synthesized audio $\hat{x}_{i}^{sp}$ pass through the shared encoder $E_{asr}$ to extract ASR-BNFs $H_{bnf}$ and $\hat{H}_{bnf}$. Then we calculate the L1 Loss to optimize the decoder and BNA layers.
\begin{equation}
        \tilde{x}_{i}^{sp} = DA_{per}(x_{i}^{sp})
\end{equation}
\begin{equation}
    \begin{split}
        \small
        H_{bnf} = E_{asr}(\tilde{x}_{i}^{sp}) \quad
        \hat{H}_{bnf} = E_{asr}(\hat{x}_{i}^{sp})
    \end{split}
\end{equation}
\begin{equation}
    \small
        \mathop{\min}_{{D(\cdot),Ada(\cdot)}}\mathcal{L}_{bnf} = \frac{1}{N}\sum_{i=1}^{N}(\parallel {\hat{H}_{bnf} - H_{bnf} \parallel_{1}})
\end{equation}
The content information in utterance should be the same after VC, only the speaker-dependent information has changed. Finally, we get relatively pure content information and have good disentangling performance with ASR-BNFs' constraint. 

\textbf{Adversarial speaker classifier}: 
We integrate a speaker classifier with a Gradient Reversal Layer (GRL) to remove the timbral information from the embedded content representation.
The expectation is that the content encoder will be trained to reduce the incorporation of speaker-specific details. The gradient is first reversed by the GRL to aim for the diminishment of speaker-specific attributes within the content embedding, prior to its backward propagation towards the content encoder. The adversarial loss is as follows:
\begin{equation}
\begin{split}   
        \small
        \mathop{\min}_{E_{c}(\cdot)Ada(\cdot)}\mathcal{L}_{adv} = \sum_{n=1}^{N}\mathbb{I}(id_{spk}==n)\log P_{spk}^{n}
\end{split}
\end{equation}
where $\mathbb{I}(\cdot)$ is the indicator function, $P_{spk}^{n}$ represents the probability that the speaker classifier's output, which indicates the probability of being classified as $spk_{n}$. 


\textbf{Timbre fusion}: 
Inspired by \cite{sanet}, we design a decoder with Speaker-Attention (SAT) block, which learns the mapping between the timbre features $F_{spk}$ and the fusion features $Z$ by slightly modifying the cross-attention mechanism. The SAT is designed as follows:
\vspace{-0.3cm}

\begin{small}
\begin{equation}
\begin{split}
        Z_{con} &= \mathbf{Con}(Z_{c},Z_{p},Z_{r})  \\
        Z = \mathbf{Con}(Softmax((&W_{q}Z_{con})(W_{k}F_{spk})^{T})W_{v}F_{spk} ,Z_{con})
\end{split}
\end{equation}
\end{small}


\noindent where $\mathbf{\small Con}(\cdot)$ means concatenation. Since it is difficult to accurately estimate speaker representations for unseen speakers, using inaccurate speaker representations as input to the decoder will lead to a mismatch between training and inference. The SAT block was designed to improve the generalization to unseen speakers compared to concatenation \cite{speechsplit1}.
Finally, the training loss of voice conversion is:
\begin{equation}
    \begin{split}
    \small
        \mathcal{L}=\mathcal{L}_{recon} + \alpha_{1}\mathcal{L}_{MI} +\alpha_{2}\mathcal{L}_{con}+\alpha_{3}\mathcal{L}_{adv}+\alpha_{4}\mathcal{L}_{bnf}
    \end{split}
\end{equation}

\normalsize
\section{Experiments}
\normalsize
\label{sec:print}

\subsection{Experiment setup}
We use the VCTK corpus \cite{vctk}, which are randomly split into 100, 3 and 6 speakers as training, validation and testing sets respectively. Each speaker has about 400 sentences, and the audio is downsampled to 16kHz.
To perform objective and subjective tests, we randomly select 50 conversion pairs to generate synthesized samples from both models. 20 listeners participated in subjective evaluation and scored the naturalness and similarity of the test audio.

In stage 1, we pre-train our content, pitch, and rhythm encoders for 50k iterations, with the learning rate set to 5e-5. Additionally, the temperature t in Eq.\ref{nce_tem} is set to 0.1. 
Due to the presence of numerous variables, we employ cosine annealing \cite{cosineannel} to adjust the learning rate and govern the gradient update step, mitigating the risk of falling to a local optimal solution. At the same time, we use learning rate warm-up \cite{warmup} to slow down the overfitting phenomenon of the mini-batch model in the initial stage and maintain the stability of the distribution.

In stage 2, The BNA and decoder are trained using a learning rate of 1e-4 for 500k iterations and we use a pre-trained WaveNet as vocoder.

\begin{table*}[t!]
\caption{Subjective and Objective Evaluation results of different methods.}
\label{tab1}
\centering
\scalebox{0.99}{
\begin{tabular}{ccccccccccc}
\toprule
\cmidrule(r){1-11}
\multirow{2.5}{*}{  \textbf{Methods}} & \multicolumn{5}{c}{\textbf{Many-to-Many VC}} & \multicolumn{5}{c}{\textbf{One-Shot VC}} \cr 
\cmidrule(lr){2-6} \cmidrule(lr){7-11} &MOS$\uparrow$ &SMOS$\uparrow$ &MCD$\downarrow$ &$\log F0$ PCC$\uparrow$ &WER$\downarrow$ &MOS $\uparrow$ &SMOS$\uparrow$ &MCD$\downarrow$ &$\log F0$ PCC$\uparrow$ &WER$\downarrow$\cr   \midrule
SpeechFlow \cite{speechsplit1} & 3.56$\pm$0.08   & 3.04$\pm$0.11  & 6.37 & 0.686 & 23.5\%   
& 2.62$\pm$0.10  & 2.45$\pm$0.09  & 8.01 & 0.544 & 31.7\%    \\
VQMIVC \cite{vqmivc}    & 3.70$\pm$0.13    & 3.61$\pm$0.09  & 5.46 &\bfseries 0.829 & 16.9\%
& 3.35$\pm$0.09   & 3.06$\pm$0.11  & 6.77 & 0.675 & 24.1\% \\
Liu et al. \cite{auto-disentangle} & 3.62$\pm$0.09   & 3.20$\pm$0.16 & 7.25 & 0.670 & 25.6\%
& 3.05$\pm$0.11   & 2.89$\pm$0.14  & 7.59 & 0.552 & 29.3\% \\
\textbf{EAD-VC(Con)} &\bfseries 3.85$\pm$0.16 &\bfseries 3.64$\pm$0.11 & 5.40 & 0.758 & 15.2\%
& 3.47$\pm$0.16 & 3.05$\pm$0.17 & 6.45 & 0.606 & 22.8\% \\
\textbf{EAD-VC} & 3.83$\pm$0.13 & 3.63$\pm$0.12 &\bfseries 5.21 & 0.793 &\bfseries 14.6\%
&\bfseries 3.52$\pm$0.14 &\bfseries 3.35$\pm$0.19 &\bfseries 6.33 &\bfseries 0.693 &\bfseries 21.5\% \\
\midrule
w/o BNA\&$L_{MI}$  &3.66$\pm$0.13   &3.38$\pm$0.17 &5.96 & 0.749 &16.4\% &3.21$\pm$0.13  &2.92$\pm$0.13 &7.12 & 0.622 &25.9\% \\
w/o $L_{con}$  &3.71$\pm$0.15 &3.46$\pm$0.14 &5.37 & 0.745 &21.1\%  &3.36$\pm$0.09  &2.98$\pm$0.18 &6.96 & 0.675 &28.6\% \\
w/o $L_{bnf}$  &3.80$\pm$0.11 &3.58$\pm$0.12 &5.23 & 0.762 &18.4\%  &3.41$\pm$0.11  &3.04$\pm$0.11 &6.51 & 0.676 &27.1\% \\
w/o $L_{spk}$  &3.75$\pm$0.09 &3.51$\pm$0.11 &5.31 & 0.737 &17.6\%  &3.37$\pm$0.12  &2.94$\pm$0.14 &6.73 & 0.648 &23.4\% \\
\bottomrule
\end{tabular}}
\vspace{-0.4cm}
\end{table*}

\subsection{Evaluation}
We compare our proposed method with other systems, including:
\begin{itemize}
\item SpeechFlow \cite{speechsplit1}, which can decompose speech into pitch, rhythm, content, and timbre by introducing three carefully designed information bottlenecks;
\item VQMIVC \cite{vqmivc}, which use vector quantization to generate content embedding and employ MI as the correlation metric to decompose speech;
\item Liu et al. \cite{auto-disentangle}, which is the first method to automatically decouple speech into different components through data enhancement and rank modules without hand-crafted features. 
\item EAD-VC: The VC method we proposed using SAT to fuse timbre; EAD-VC(Con) is a method of concatenating timbre in the channel dimension which is the same as the previous work \cite{speechsplit1}.
\end{itemize}

\subsection{Subjective evaluation results}
Listeners utilize subjective evaluation to gauge the naturalness of speech and the similarity of the speaker's voice in the converted speeches, which are produced by a variety of models.
We use Mean Opinion Score (MOS) to describe the naturalness of various model outputs, and Speaker Similarity MOS (SMOS) to evaluate the likeness between the converted audio and the intended audio with 95\% confidence intervals, including timbre and prosody. 20 listeners (10 males and 10 females) are requested to assign scores within a range of 1 to 5 points respectively.

As depicted in Table \ref{tab1}, EAD-VC(Con) enhances the similarity to target speakers and achieves higher speech naturalness than other systems in many-to-many VC. In one-shot VC scenario, the method of directly concatenating content, rhythm, pitch and timbre in the channel dimension will significantly reduce the naturalness and similarity of converted speech. 
The Timbre fusion method of timbre components concatenation is insufficient to simulate the timbres of various unseen speakers in real life. 
However, the performance drop of EAD-VC with SAT is smaller than that of other models including EAD-VC(Con). SMOS metrics in one-shot VC is improved by 0.3 on EAD-VC verifies that SAT can enhance the speaker similarity and the model generalization in the face of unseeen speakers.

\subsection{Objective evaluation results}
As shown in Table \ref{tab1}, we calculate mel-cepstrum distortion (MCD) and word error rate (WER) for converted audio. To evaluate pitch variations of the converted audio, the Pearson correlation coefficient (PCC) between F0 of the source and the converted audio is calculated. 

The EAD-VC and EAD-VC(Con) have the lowest WER among the various methods in preserving linguistic content. This shows that our method is very robust.
At the same time, the lowest MCD among all methods demonstrates that our method improves the speaker similarity for less distortion between converted audio and target audio. 
By controlling the pitch variations, we can achieve high F0 consistency in the audio, as demonstrated by the $F0$-PCC obtained from EAD-VC, which is higher than other methods except \cite{vqmivc} in many-to-many VC. It proves that the converted audio generated by EAD-VC(Con) has a pitch contour that is more similar to the target audio. The $F0$-PCC of \cite{vqmivc} is higher than EAD-VC(Con), we attribute this to the fact that the pitch embedding extracted from source speech in \cite{vqmivc} is an external supervised label that is fed directly to the decoder without passing through the pitch encoder. However, in the face of unseen speakers in one-shot VC scenarios, \cite{vqmivc} will degrade model performance more than EAD-VC which can disentangle pitch information from speech. It proves the advantages of SAT in One-shot VC compared to concatenation.

Furthermore, an additional evaluation involving a fake speech detection test is conducted in one-shot VC condition.
Using Resemblyzer toolkit\footnote{Resemblyzer toolkit: https://github.com/resemble-ai/Resemblyzer}, we randomly select 6 of the 10 real audios as groundtruth.
The remaining four real audio samples along with the converted audio from diverse models will serve as the basis for evaluating timbral resemblance.

\begin{figure}[htb]
\vspace{-0.3cm}
    \centering
    \subfigure[F2M]{
        \label{same-gender conversion 1}
        \begin{minipage}[b]{0.44\linewidth}
            \includegraphics[width=1\textwidth]{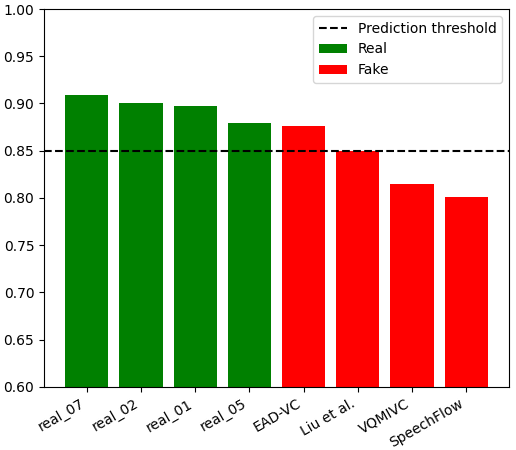}
        \end{minipage}
    } 
    \subfigure[M2F]{
         \label{cross-gender conversion 1}
        \begin{minipage}[b]{0.44\linewidth}
            \includegraphics[width=1\textwidth]{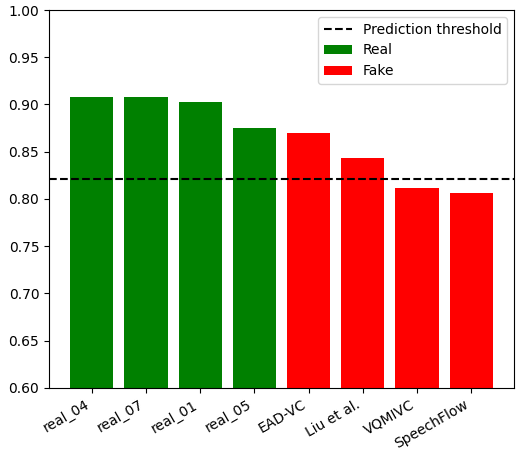}
        \end{minipage}
    }
    \vspace{-0.2cm}
    \caption{Scores pertaining to VC are indicated. F denotes Female, and M denotes Male. Different models are represented on the x-axis, while prediction scores are represented on the y-axis.}
    \label{fake}
\end{figure}

As shown in Figure \ref{fake}, the scores for real audios are indicated by the green clusters, whereas the red clusters correspond to the scores of the synthesized audios. The dashed line signifies the predictive threshold; any score surpassing this line is categorized as belonging to a real audio.
Our proposed EAD-VC outperforms the other models on F2M and M2F VC by coming to highest scores over the dash-line among fake samples.

\subsection{Generalization to unseen speaker}

\begin{figure}[h]
    \centering
    \vspace{-0.3cm}
    \includegraphics[width=9.0cm]{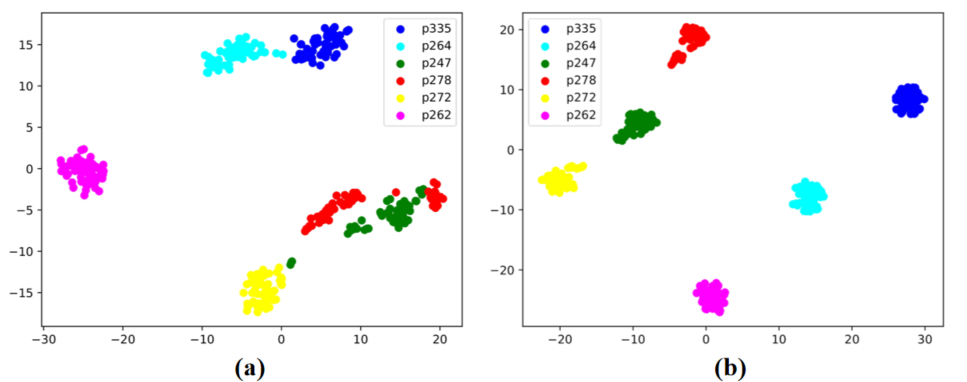}
    \vspace{-0.5cm}
    \caption{Timbre embedding visualization on One-Shot VC. (a) Liu et al. \cite{auto-disentangle}; (b) EAD-VC.}
    \label{visual}
\end{figure}

Fig.\ref{visual} illustrates the different timbre embeddings visualized by the tSNE method. To prove the effectiveness of our method in enhancing speech representation auto-disentangling, we choose Liu et al. \cite{auto-disentangle} as the baseline. 
The difference between the two is that EAD-VC uses timbre embeddings extracted from speaker encoder which jointly trained in the model and fused with other speech components, while the timbre embedding extracted from the pre-trained speaker encoder in \cite{auto-disentangle} is directly concated with other speech components in the inference stage.

As shown in Fig.\ref{visual}(b), the better the clustering effect of embedding, the stronger the disentangling ability of VC model \cite{Multi-target-VC}. We achieves significant disentangling effects in one-shot VC. It demonstrates that EAD-VC can produce more clustered timbre representations on both VC and one-shot VC and disentangle timbre information from speaker information. 
When we need to fine-tune the VC model on new unseen speakers, the encoding part only needs to update the parameters of the adaptor connected in series behind the disentangled encoder. 
Compared with other one-shot VC which require updating all model parameters, our method can reduce memory usage during training and improve computational efficiency.

\subsection{Ablation study results}
To validate the effect of our proposed method, we conduct ablation experiments. As shown in Table \ref{tab1}, when the model is trained without the $L_{MI}$ of bottleneck adaptor and ($L_{con}$, $L_{bnf}$, $L_{spk}$) of TGC module, both the quality and similarity scores drop when removing them. It still outperforms most of Liu et al., which demonstrates the efficiency of BNA\&MI and TGC in improving speech naturalness and similarity.

When BNA\&$L_{MI}$ is removed, our method failed in almost all metrics in VC task, especially the MOS and SMOS. It means the $L_{MI}$ is important in removing information overlap between decoupled speech components.
What's more, we observe that ASR performance (WER) and MCD drop significantly without using $L_{con}$, 
since the converted voice is compromised by undesired content information entangled with speaker representations.
The existence of $L_{con}$ can greatly reduce this timbre leakage problem. It also outperforms the model lacking $L_{bnf}$ in terms of WER, which indicates the effectiveness of speech2content module in content inconsistency before and after voice conversion.
Furthermore, it's shown that the conversion quality of F0 is significantly degraded without the $L_{spk}$ of the adversarial speaker classifier.


To assess the effectiveness of automatically disentangling pitch and rhythm through data augmentation and self-supervised learning, we employed the real pitch and rhythm extracted from speech as the label for supervised learning. We adopt this direct prediction approach to disentangle \textbf{P}itch (F0) and \textbf{R}hythm and compare it with our method in Table \ref{tab2}.

\begin{table}[t]
  \caption{Ablation Study of P\&R Decoupling Effectiveness.}
  \label{tab2}
  \centering
  \resizebox{0.40\textwidth}{!}
  {
  \begin{tabular}{cccc}
    \toprule
    \cmidrule(r){1-4}
     \textbf{Guidance} &\textbf{MOS}$\uparrow$ &\textbf{SMOS}$\uparrow$ &\textbf{MCD}$\downarrow$     \\
    \midrule
    with \textbf{F0}  & 3.87$\pm$0.09 & 3.75$\pm$0.09 & 5.15      \\
    with \textbf{Rhythm} & 3.67$\pm$0.11 & 3.41$\pm$0.13 & 6.11     \\
    with \textbf{F0}\&\textbf{Rhythm} & 3.72$\pm$0.08 & 3.52$\pm$0.11 & 5.76      \\  
    \midrule
    \textbf{Ours} & 3.83$\pm$0.13 & 3.63$\pm$0.12 & 5.21      \\ 
    \bottomrule
  \end{tabular}
  }
  \vspace{-0.5cm}
\end{table}

When incorporating only real pitch as a guide during the training process (similar to VQMIVC), results show that supervised training using real pitch (F0) as labels can improve naturalness and similarity to a certain extent.
What's more, the pre-train rhythm encoder \cite{speechsplit2} is used to extract rhythm embedding from speech as a guide for supervised learning.
It proves that extracting rhythm information from speech is a challenging task to a certain extent as the disentangling effect drops significantly with rhythm guidance as shown in Table \ref{tab2}. Our method can effectively disentangle rhythm from the latent space and does not require manual extraction of rhythm as a guide, which proves the feasibility of our method of disentangling pitch and rhythm.

\subsection{Conversion rate}
To demonstrate that our approach can enhance speech auto-disentanglement, we evaluate the conversion rate of the baseline \cite{auto-disentangle} and EAD-VC. The conversion rate can serve as an indicator of the effectiveness of the disentangling process, which is followed by \cite{speechsplit1}. 
\vspace{-0.25cm}
\begin{figure}[h]
    \centering
    \vspace{-0.1cm}
    \includegraphics[width=8.0cm]{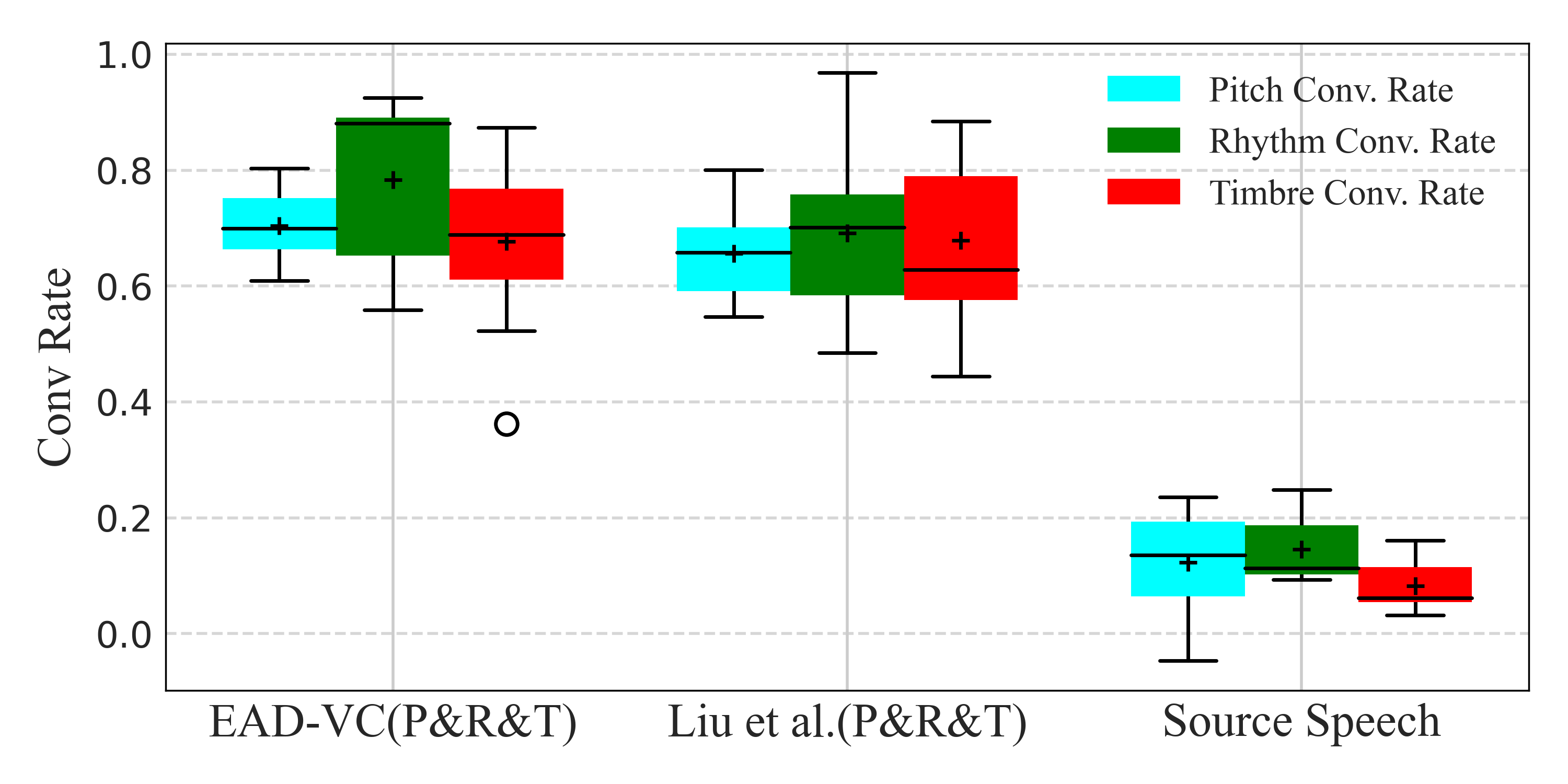}
    \vspace{-0.1cm}
    \caption{Subjective conversion rate Evaluation. Each group encompasses three distinct subsections, which correspond to the conversion rates of pitch, rhythm, and timbre.} 
    \vspace{-0.25cm}
    \label{convert_rate}
\end{figure}

As shown in Fig.\ref{convert_rate}, the conversion rate of our model exceeds the baseline in Pitch and Timbre.
The findings indicate that while our disentangled encoders are structurally identical to the baseline, training them with BNA\&$L_{MI}$ and TGC modules enhances our model's effectiveness in extracting disentangled pitch, rhythm, and timbre information. 


\section{Conclusions}
\label{sec:prior}
In this paper, we introduce an innovative VC model with the designed IFUB estimator and joint text-guided consistent learning that can achieve SSL-based speech representation disentanglement.
We use MI minimization with IFUB estimator to enhance the ability to speech auto-disentanglement into four components without the need for manual input of hand-crafted features and eliminate information overlap between components.
Moreover, we use TGC module to solve timbre leakage and avoid content inconsistencies after VC. The experiments demonstrate that the EAD-VC which we have introduced, is capable of delivering superior disentanglement results and generate more authentic and natural speech.

\section{Acknowledgement}
Supported by the Key Research and Development Program of Guangdong Province (grant No. 2021B0101400003) and Corresponding author is Xulong Zhang (zhangxulong@ieee.org).

\bibliographystyle{IEEEtran.bst}
\bibliography{EAD-VC.bib}

\end{document}